\documentstyle[12pt]{article}

\begin{document}

\setcounter{page}{100}

\centerline{\large {\bf The Mandelstam--Leibbrandt prescription and}}
\centerline{\large {\bf the Discretized Light Front Quantization.}}
\vskip.3in
\centerline{ Roberto Soldati}
\centerline{ Dipartimento di Fisica "A. Righi", Universit\`a di Bologna}
\centerline{ via Irnerio 46, 40126 Bologna, Italy}

\vskip.6 in
\centerline{\bf Abstract}
\vskip.1in
It is shown that the quantization of the unphysical degrees of freedom, which
leads to the Mandelstam--Leibbrandt prescription for the infrared spurious
singularities in the continuum light cone gauge, does indeed suggest some
quite natural recipe to treat the zero modes in the Discretized Light Front
Quantization of gauge theories.
\vskip.1in

\vskip.3in
{\bf 1. Introduction}
\vskip.1in

Light Front Dynamics (LFD) of field theories, in which $x^+=x^0+x^3$
plays the role of the evolution parameter, has many appealing and
useful features$^1$. Among them, the maybe most important one concerning the
quantum theory, is the occurrence of a nonperturbative vacuum simpler
than in the ordinary time formulation.
In the case of gauge theories, the LFD leads to the light--cone gauge as the
most convenient choice for the subsidiary condition. However, owing to the need
of defining the inverse of $\partial_-\equiv {\partial\over \partial x^-}$,
$x^-=x^0-x^3$, the difficult problem arises of a consistent
handling for the infrared spurious singularities.\par
Since the very early attempts$^{2,3}$ to deal with the above matter,
the attitude
was the following: the zero modes, associated to $\partial_-$, are eliminated
assuming suitable boundary conditions for all the fields at $x^-=\pm\infty$
and,
consequently, the spurious infrared singularities are defined, in the
momentum space, through the Cauchy Principal Value (CPV) prescription (or some
equivalent to it). It turns out that the ensuing Feynman perturbation theory
does not fulfil any power counting criterion and eventually leads to
inconsistent results, even at one loop$^4$ as in the SUSY N=4 model.
As a consequence, the above mentioned philosophy is ruled out by explicit
perturbative calculations. \par
In order to restore the
agreement between light-cone gauge and covariant gauge
perturbative calculations, in the SUSY N=4 model, S. Mandelstam proposed to
define$^5$ the spurious infrared singularities as follows:
$$
{1\over [k_-]}\equiv
\lim_{\epsilon\to 0^+} {1\over k_-+i\epsilon sgnk_+}\quad,
\eqno (1.1)
$$
where the limit is understood in the sense of distribution (an alternative, but
equivalent, form$^6$ has been proposed by G. Leibbrandt).
Shortly afterwards it has been shown
that the Mandelstam--Leibbrandt (ML) prescription (1.1)
originates from canonical
{\bf equal time} quantization$^7$
and, later on, that
the corresponding Feynman perturbation theory lies on the same firm ground
as in the covariant gauges. As a matter of fact, the ML
prescription fulfils generalized power counting, it allows the Wick rotation
in the Feynman integrals and, very remarkably, it leads to perturbative
renormalizability and unitarity$^{7,8}$, once in the effective action
some non-local and non-covariant
counterterms are introduced,
which are {\bf completely determined}
to all order in the loop expansion.

\vskip.3in
{\bf 2. The continuum Light Front formulation}
\vskip.1in

As previously mentioned, the ML prescription naturally emerges from the
ordinary equal time canonical quantization. Very recently, it has been
shown$^{9}$ that actually the ML form of the propagator can be obtained
from a Light Front formulation, provided some zero modes are properly
taken into account and suitably quantized. Let me briefly recall the main
points of the derivation.\par
The lagrangean density of the free radiation field in the light--cone gauge
is given by
$$
{\cal L}=-{1\over 4}F_{\mu\nu}F^{\mu\nu}+\partial_k\lambda\partial_k A^+
\quad ,
\eqno(2.1)
$$
with $x_\perp=(x^1,x^2)$, $j,k,... = 1,2$, $A^\pm = A^0\pm A^3$, the evolution
being along $x^+$. The subsidiary condition $A^+=0$ immediately follows,
if  the boundary condition $A^+(x^\pm,x_\perp)\to 0$, when $|x_\perp|\to 0$,
is assumed and the equations of motion read
$$\partial_-^2A_+ + \partial_-\partial_k A^k = 0\quad ,
\eqno(2.2a)
$$
$$
(4\partial_+\partial_- - \partial_\perp^2) A_k -
\partial_k (\partial_- A_+ + \partial_j A^j) = 0\quad ,
\eqno(2.2b)
$$
$$
2\partial_+\partial_- A_+ - \partial_\perp^2 A_+
- 2\partial_+\partial_k A^k = 2\partial_\perp^2 \lambda\quad ,
\eqno(2.2c)
$$
leading to $\partial_-\lambda = 0$.
If we impose the boundary condition $\lambda\to 0$ when $x^-\to \pm\infty$,
then $\lambda\equiv 0$ and no zero modes are present.
However, as previously emphasized,
this eventually yields the CPV prescription for the spurious
singularity in the Feynman propagator and, therefore, to the inconsistent
perturbation theory. To be consistent we have to keep $\lambda=\lambda
(x^+,x_\perp)\not= 0$, which has to be correctly determined within the
Light Front formalism. To this aim let us define
$$
A_k (x)\equiv T_k (x) + {\partial_k\over \partial_\perp^2}\varphi
(x^+,x_\perp)\quad ;
\eqno(2.3)
$$
from the equations of motion we obtain
$$
(4\partial_-\partial_+ - \partial_\perp^2)T_k=0\quad ,\eqno(2.4a)
$$
$$
A_+ = \partial_-^{-1}\partial_k T_k - 2
(\lambda + {\partial_+\over \partial_\perp^2} \varphi) =
{4\partial_+\over \partial_\perp^2}\partial_k T_k - 2
(\lambda + {\partial_+\over \partial_\perp^2} \varphi)\quad ,
\eqno(2.4b)
$$
since we are working with {\it on shell} free fields $T_k$.\par
Now, in order to find some Light Front operator algebra {\bf isomorphic}
to the canonical equal time operator algebra, we have to impose
($\tilde v\equiv (v^-,v_\perp$))
$$
[T^j (x),\partial^+ T^k (y)]_{x^+=y^+} = i \delta^{jk} \delta^{(3)}
(\tilde x -\tilde y)\quad ;
\eqno(2.5)
$$
$$
[\varphi (\tilde x),\lambda (\tilde y)] =
i \delta (x^+ - y^+ )\delta^{(2)}(x_\perp - y_\perp )\quad ;
\eqno(2.6)
$$
$$
[T^k (x), \varphi (\tilde y)]=[T^k (x), \lambda (\tilde y)]=[\varphi
(\tilde x),
\varphi (\tilde y)]=[\lambda (\tilde x),\lambda (\tilde y)]=0\quad .
\eqno(2.7)
$$
Some key remarks are in place concerning the above operator algebra:
namely,\par
i)\quad the commutator $[\varphi (\tilde x),\lambda (\tilde y)]_{x^+=y^+}$
does not make sense. This means that it is not possible to simultaneously
specify all the fields on the same ``initial'' hyperplane $x^+=constant$,
but one has to specify the zero mode commutators for different (not
coincident) Light Front ``times''. A related feature is that the canonical
Light Front Hamiltonian $P_+$ does not provide the evolution of the zero
mode fields (see also below).\par
ii)\quad The space of the state vectors is an indefinite metric linear
space, as we already know from canonical equal time quantization$^7$.\par
iii)\quad In the present free case, we have that all the components of
$A_\mu (x)$ do indeed vanish when $x^-\to\pm\infty$; consequently the
theory is allowed to be formulated on a compact domain along $x^-$ in the
presence of periodic boundary conditions.

\vskip.3in
{\bf 3. Discretized Light Front Quantization}
\vskip.1in
The Discretized Light Front Quantization
(DLFQ) has been proposed$^{10}$ to provide an infrared cut--off for
the spurious
singularities and some alternative non--perturbative computer algorithm
other than Euclidean lattice QCD;
moreover one can easily appreciate the non trivial
features associated with the onset of the zero modes. Let us define our
theory on the hypercylinder $\Omega^- = \{x^\mu|
x^+\in {\hbox{\bf R}}, x_\perp\in
{\hbox{\bf R}}^2; x^-\in [-L,L]\}$ and impose to $A_\mu$
periodic boundary conditions: namely,
$$
A_\mu(x)=A_\mu^\circ (x^+,x_\perp)+
\sum_{n\not=0}A_\mu^n (x^+,x_\perp) \exp\left\{i{\pi n\over L}x^-\right\}
\quad ,
\eqno(3.1)
$$
the zero modes $A_\mu^{n=0}\equiv A_\mu^\circ$ being now independent fields
in the LFD.\par
Let us first discuss the free radiation field. The normal mode sector,
$n\not= 0$, can be treated according to the usual Light Front formulation,
since the derivative $\partial_-$ can be inverted as
$$
(\partial_-^{-1}*\Phi)(x)=\sum_{n\not=0}{L\over i\pi n}\Phi^n(x^+,x_\perp)
\exp\left\{i{\pi n\over L}x^-\right\}\quad ,
\eqno(3.2)
$$
where $\Phi$ is any of the normal field components.
Among the zero modes, the component $A_-^\circ$ is gauge invariant, since
the infinitesimal gauge transformation $\delta A^+(x)=\partial_-\Lambda(x)$
involves a periodic function $\Lambda$. The lagrangean density for the zero
modes can be written as
$$
{\cal L}_{ZM}={1\over 2}(\partial A_-^\circ)^2 - {1\over 2}(F_{12}^\circ)^2
-F_{+k}^\circ\partial_k A_-^\circ - A_-^\circ\partial_\perp^2\lambda\quad ,
\eqno(3.3)
$$
which is singular and leads to the primary first class constraints
$$
\pi_-^\circ\approx 0\quad ;\quad
\rho_k^\circ\equiv \pi_k^\circ - \partial_k A_-^\circ\approx 0\quad .
\eqno(3.4)
$$
It should be stressed that, since the constraints $\rho_k^\circ\approx 0$
are first
class, at variance with the corresponding ones in the normal mode sector
which are second class, there is an additional ``transverse'' gauge
invariance in the zero mode sector. As a matter of fact,
the equations of motion for the zero modes: namely,
$$
\partial_\perp^2 A_-^\circ = 0\quad \Longrightarrow\quad A_-^\circ=0\quad ;
\eqno(3.5a)
$$
$$
\partial_\perp^2 A_+^\circ + \partial_-\partial_k A_k^\circ +
\partial_\perp^2 \lambda = 0\quad ;
\eqno(3.5b)
$$
$$
(\partial_\perp^2\delta_{jk} - \partial_j\partial_k) A_k^\circ = 0\quad ,
\eqno(3.5c)
$$
do indeed explicitly exhibit the ``transverse'' gauge invariance
(notice that the canonical Light Front zero mode Hamiltonian $P_+^\circ$
is weakly vanishing, thereby preventing the ordinary $x^+$ evolution
for the zero modes, as already mentioned). There are
infinitely many ways, of course, to remove the above
residual local gauge freedom$^{11}$ . However, at the quantum level,
this entails the pathological CPV propagator in the continuum limit
$L\to\infty$. On the contrary,
the requirement of a smooth transition
to the consistent continuum formulation
actually suggests to keep that freedom and,
instead of removing the gauge degrees of freedom, one is led to
impose the following zero mode commutation relations: namely,
$$
[\varphi (x^+,x_\perp),\lambda (y^+,y_\perp)]= i \delta (x^+-y^+)
\delta^{(2)} (x_\perp - y_\perp)\quad ,
\eqno(3.6)
$$
where $A_j^\circ (x^+,x_\perp)=\partial_j (\partial_\perp^2)^{-1}
\varphi (x^+,x_\perp)$,
in perfect analogy with eq.~(2.6).
The above recipe ensures that, in the continuum limit, the correct
ML quantization scheme is indeed recovered.\par
The interaction with spinorial matter requires the introduction of the
two Light Front components of the Dirac field $\psi_\pm = {1\over 2}
\gamma^0\gamma^\pm \psi$, satisfying antiperiodic boundary conditions
($i.e.$ no fermion zero modes), in such a way to get a periodic fermion
current $J_\mu(x)$.
 Among the Maxwell equations for the zero
modes we have
$$\partial_\perp^2 A_-^\circ + J_-^\circ =0\quad ,
\eqno(3.7)
$$
involving the component of the potential to be gauged away.
At first sight, eq.~(3.7) seems to prevent$^{11}$ the usual light--cone
subsidiary condition $A_-^\circ =0$; nonetheless, one can fulfil
the {\bf strong} light--cone gauge, provided one requires $J_-^\circ= 0$. This
constraint on the fermion field component $\psi_-$, which is not the
independent one,
is quite acceptable in the discretized ($i.e.$ regularized)
formulation. As a matter of fact, the physical
fermion current in the continuum can not contain zero modes, if
we ask the
charge $Q_-$ to be finite after the removal of the infrared
regularization along $x^-$.
Once again, in order to solve the dynamics of the gauge potential
zero modes, we still do not eliminate the redundant degrees of freedom
and, after setting
$A_k^\circ = T_k^\circ + \partial_k (\partial_\perp^2)^{-1} \varphi$,
we impose the commutation relations (3.6), the quantities $T_k^\circ$
being determined by equations of motion in the zero mode sector. In this
way the continuum limit $L\to\infty$ does reproduce, step by step in
perturbation theory, the consistent formulation including the ML zero
modes $\lambda$ and $\varphi$. It should be stressed once more that
to drop the latter ones$^{11}$ is not a safe procedure,
in order to be guaranteed
about covariance, causality, perturbative
gauge invariant renormalizability and unitarity, when the infrared spurious
regularization is removed, namely in the physical continuum limit
$L\to\infty$.

\vskip.3in
{\bf References}
\vskip.1in

1. P.A.M. Dirac, Rev. Mod. Phys. {\bf 21}, 392 (1949).\par
2. E. Tomboulis, Phys. Rev. D {\bf 8}, 2736 (1973).\par
3. J.H. Ten Eyck and F. Rohrlich, Phys. Rev. D {\bf 9}, 2237 (1974).\par
4. D.M. Capper, J.J. Dulwich and M.J. Litvak,
        Nucl. Phys. {\bf B 241}, 463 (1984).\par
5. S. Mandelstam, Nucl. Phys. {\bf B 213}, 149 (1983).\par
6. G. Leibbrandt, Phys. Rev. D {\bf 29}, 1699 (1984).\par
7. A. Bassetto, G. Nardelli and R. Soldati,\par\quad
   ``{\it Yang--Mills theories in algebraic noncovariant gauges:\par\quad
   Quantization and Renormalization}'',
   World Scientific, Singapore (1991).\par
8. C. Becchi, in ``{\it Physical and Nonstandard Gauges}'',\par\quad
   Springer-Verlag, Berlin, 177 (1990).\par
9. G. McCartor and D.G. Robertson, preprint SMUHEP/93-20 (1993).\par
10. H.C. Pauli and S.J. Brodsky, Phys. Rev. D {\bf 35}, 1493 (1985).\par
11. A.C. Kalloniatis and H.C. Pauli, Z. Phys. C {\bf 60}, 255 (1993);\par
    \qquad
    preprint MPIH-V2 (1994).
\end{document}